\documentstyle [12pt,epsf]{article}

\input{epsf}
\begin{document}
\begin{titlepage}
\begin{center}

{\Large \bf Zero-point energy of massless scalar fields in 
the presence of soft and semihard boundaries in D dimensions}\\
\vspace{.3in}
{\large\em F. $Caruso^{a,b,}$\footnote[1]{e-mail: caruso@lafex.cbpf.br},
           R. De $Paola^{a,}$\footnote[2]{e-mail: rpaola@lafex.cbpf.br}
      and N.F. $Svaiter^{a,}$\footnote[3]{e-mail: nfuxsvai@lafex.cbpf.br}}\\
\vspace{.3in}

$^{a}$Centro Brasileiro de Pesquisas F\'\i sicas - CBPF\\ Rua Dr. Xavier
 Sigaud 150, Rio de Janeiro, RJ, 22290-180, Brazil\\
\vspace{.4cm}
$^{b}$Instituto de F\'\i sica - Universidade do Estado do Rio de Janeiro \\
 Rua S\~ao Francisco Xavier 524, Rio de Janeiro, RJ, 20559-900, Brazil\\
 
\end{center}
\vspace{0.5in}

\subsection*{Abstract}

The renormalized energy density of a massless scalar field defined in a 
D-dimensional flat spacetime is computed in the presence of ``soft" and 
``semihard" boundaries, modeled by some smoothly increasing potential 
functions. The sign of the renormalized energy densities for these different 
confining situations is investigated. The dependence of this energy on $D$
for the cases of ``hard" and ``soft/semihard" boundaries are compared.

\hspace{-.6cm}PACS categories: 03.70.+k, 12.20.Ds, 04.62.+v.

\end{titlepage}
\newpage
\baselineskip .37in
\section {Introduction}

In many situations in Quantum Field Theory it is assumed that the fields are 
defined in a region limited by some finite ``classical" cavity and submitted 
to some particular boundary condition. ``Classical" means here that the 
boundary which confines the fields has a very precise spatial 
location and a well defined geometrical shape (``hard" boundary). In the 
majority of the papers discussing the Casimir effect these `` hard" boundaries 
are assumed, although they are unquestionably an idealization. 
In this scenario, one can argue in what extension does the accumulated 
experience in the Casimir effect depend on the assumption of this kind 
of boundaries. In some papers these conditions 
have been relaxed --- the most recent ones quoted in Refs. [1-5].

In this paper we will deepen into the investigation on how the Casimir 
energy of a massless scalar field, defined in a D-dimensional flat spacetime,
depends on the boundary conditions and on the dimensionality. 
For this purpose, three different kinds of confining boundaries are considered:
``hard", ``soft" and also ``semihard" ones. The meaning of this 
terminology will be clarified latter.

The problem of determining the expectation value of a physical observable is 
related to the question: how to implement a renormalization scheme in a given 
situation? In 1948 Casimir presented a scheme to obtain a finite result from 
the divergent zero-point energy of the electromagnetic field \cite{cas}. 
Although formally divergent, the difference between the vacuum energy of 
different physical configurations can be finite. If one of these configurations
is assumed to have a zero vacuum energy, then the difference of the vacuum
energy of both configurations is the renormalized one. Therefore the formal 
definition of the Casimir energy is
\begin{equation}
 E_{ren}(\partial\Omega)=E_{0}(\partial\Omega)-E_{0}(0),
\label{cas}
\end{equation}
where $E_{0}(\partial\Omega)$ and $E_{0}(0)$ are, respectively,
the zero-point energies in the presence and in the absence of  
boundaries. In the case of scalar fields, Casimir's approach can be 
summarized in the following steps: a complete set of mode solutions of the 
Klein-Gordon equation satisfying an appropriate boundary condition and the 
respective eigenfrequencies are found; the divergent zero-point energy is 
regularized by the introduction of an ultraviolet cut-off and, finally, the 
polar part of the regularized energy is removed using a renormalization 
procedure.

It is well known that there are two quantities which might be expected to 
correspond to the total renormalized energy of quantum fields \cite{deu}. 
The first is called the mode sum energy $<E>_{ren}^{mode}$,
\begin{equation}
 <E>_{ren}^{mode}=\int^{\infty}_{0}d\omega\frac{1}{2}\omega
 (N(\omega)-N_{0}(\omega)),
\label{mod}
\end{equation}
where $\frac{1}{2}\omega$ is the zero-point energy for each mode, 
$N(\omega)d\omega$ is the number of modes with frequencies between $\omega$ 
and $\omega+d\omega$ in the presence of boundaries and $N_{0}(\omega)d\omega$ 
is the corresponding quantity evaluated in empty space. Eq. (\ref{mod}) gives 
the renormalized sum of the zero-point energy of each 
mode. The second one is the volume integral of the renormalized energy density, 
$<E>_{ren}^{vol}$, obtained by the Green's function method \cite{bro}. 
In the latter method, in order to calculate the renormalized energy for 
any field, a certain second order differential operator is applied
to the renormalized Green's function $G_{ren}(x,x')$, {\it i.e.},
\begin{equation}
 <T_{\mu\nu}>_{ren}=\lim\limits_{x\rightarrow x'} D_{\mu\nu}
 \left[G(x,x')-G_{0}(x,x')\right], 
\label{bro}
\end{equation}
where $G(x,x')$ is the Green's function in the presence of the boundary 
$(\partial\Omega)$ and $G_{0}(x,x')$ is the Green's function in the absence of 
boundaries. Deutsch and Candelas \cite{deu} refer to the quantity between the 
brackets as the renormalized Green's function, since both Green's 
functions give rise to the same ultraviolet singularity structure 
(as $x\rightarrow x')$. If $x$ belongs to the boundary $(\partial\Omega)$ the 
renormalized stress-tensor $<T_{\mu\nu}>_{ren}$ can diverge as one gets close 
to this surface. However, as was stressed by these authors, the above argument 
is not a proof that the renormalized stress-tensor $<T_{\mu\nu}>_{ren}$ 
will diverge as we get closer to $(\partial\Omega)$, but 
it suggests that if the renormalized stress tensor is bounded near 
$(\partial\Omega)$ it means that a delicate cancellation must occur. In the 
case of a perfectly conducting spherical shell in the presence of an
electromagnetic field both inside and outside the cavity there is a 
cancellation between the TE and TM modes, giving rise to a finite energy 
density even on the boundary \cite{DeRaad}.

It is important to point out that, for the minimally coupled 
scalar field, such cancellation does not occur, which renders the concept of 
the renormalized vacuum energy density $<T_{00}>_{ren}$ ambiguous. However it 
is well known that the total renormalized vacuum energy associated with a 
minimally coupled scalar field obtained by the sum of modes method,  
$<E>_{ren}^{mode}$, must be equal to that of the conformally coupled case, 
since both fields satisfy the same wave equation and have the same density of 
states. Nevertheless, the total renormalized vacuum energies obtained from 
the Green's function method, $<E>_{ren}^{vol}$, for the minimal and conformal 
scalar fields, are different. Actually, $<E>_{ren}^{mode}$ is found to be 
divergent. Which of these quantities, $<E>_{ren}^{mode}$ or $<E>_{ren}^{vol}$, 
therefore, is the ``physical" renormalized energy of a minimally coupled 
scalar field? In the bag model this problem is also present \cite{cho}. 
Using the Green's function method, Bender and Hays \cite{hays} obtained a 
quadratic divergence for minimal scalar fields confined in the interior of 
the bag. Also Milton, investigating the zero-point energy of vector 
fields (gluons) confined in a spherical bag \cite{mil1},  
obtained the same kind of quadratic divergence.

It has often been suggested by many authors that a 
full quantum mechanical treatment of boundary conditions can solve the 
above mentioned problem. Recently, Ford and Svaiter have confirmed
these especulations \cite{ford} assuming fluctuating boundaries. By 
considering confining plates as quantum objects with a 
position probability distribution $|\psi(q)|^{2}$, it was shown that this 
approach is able to remove the discrepancy between $<E>_{ren}^{mode}$ and 
$<E>_{ren}^{vol}$ for the minimally coupled scalar field, solving a long 
standing paradox concerning the renormalized energy of the minimally and 
conformally coupled scalar fields.

There are many other different approaches in order to relax the classical 
boundary conditions. A long time ago, investigating the bag model, some 
authors discussed quantum corrections to this model by quantizing fluctuations 
around the clasical bag solution \cite{rebi}. Working in the same 
direction, Creutz \cite{creu} studied the effects of considering 
different bag configurations using the path integral approach in a theory
with different massive scalar fields inside and outside the bag. 
Golestanian and Kardar \cite{gole} also use a path integral approach
to investigate the problem of perfectly reflecting cavities that undergo an 
arbitrary dynamical deformation. These authors were able to calculate the 
behavior of the mechanical response function ({\it i.e.}, the ratio between 
the induced force and the deformation field in the linear regime). Some 
authors, on the other hand, employed a simpler alternative approach, which 
allows one to deal with more general physical situations than the ``hard"  
classical boundary conditions currently used in the literature. They
imagine a confining ``soft" boundary as modeled by a given smoothly increasing 
potential function representing some distribution of matter which interacts 
with the quantum field  \cite{act,alb}. Using this approach, it is possible 
to recover ``hard" boundaries in some limit. This point will be
clarified latter.

The aim of this paper is to discuss the Casimir effect for  
massless scalar fields subjected not only to ``hard" boundaries in four 
dimensional spacetime but also to ``soft" and ``semihard" ones 
in a general D-dimensional flat spacetime. A classic question to be analysed 
is what actually determines the attractive or repulsive nature of the Casimir 
force. As it is well known, the sign of the Casimir energy may 
depend on the type of boundary conditions, on the ratio of the finite 
characteristic lenghts of the cavity and on many other geometrical and
topological features \cite{luk,ambj}. Recently, 
the sign of the Casimir energy was discussed in Ref. \cite{cor} 
where some results obtained in Ref. \cite{car} were generalized, but still 
assuming only Dirichlet b.c. It is our purpose here to address the 
question of the sign of the Casimir energy in these ``new" confining 
situations in a D-dimensional flat spacetime.

The paper is organized in the following way. In Section II we review the 
most simple example of the Casimir energy dependence on the ratio between 
the characteristic lenghts of a ``classical" cavity. In Section III we analyse 
the Casimir effect for a minimally coupled scalar field in a 
D-dimensional spacetime in the presence of ``hard" boundaries. In Section IV
we investigate the Casimir effect in the presence of ``soft" 
and ``semihard" boundaries. In Section V we consider the situation 
of more than one confining potential being imposed onto the field. 
Conclusions are given in Section VI. Throughout this paper 
we use $\hbar=c=k_{B}=1$.

\section{ The Casimir energy in a two-dimensional classical box}\

In order to get some insight on the problem of renormalized quantities 
confined in compact domains, we review, in this Section, a well 
known example. The most simple situation that we can imagine in which
the vacuum energy is dependent on the ratio of characteristic lengths
of a cavity is that of a minimally coupled scalar field satisfying 
classical boundary conditions in a $3D$ spacetime.  

Let us consider a free massless scalar field confined in a $ 2D $ rectangular
box satisfying Dirichlet boundary conditions. Although the presence of the
corners are unphysical features, the model was also used by Peterson,
Hansson and Johnson \cite{pet} in the study of loop diagrams of a confined
scalar field in boxes, and it is suitable for our purposes.

A free real massless scalar field $\varphi(x)$ defined in a flat spacetime
must satisfy the homogeneous Klein-Gordon equation. If we restrict the field 
to the interior of the box, the field modes are denumerable and the positive 
and negative frequency parts form a complete orthonormal set. The renormalized 
energy can be obtained after suitable regularization and renormalization 
procedure of the infinite sum of the zero-point energy of each field mode. 
Because there is no difference between the density of modes of the minimal 
and of the conformal scalar fields, the example 
below covers both situations. In 
this Section we will follow the procedure of Ref. \cite{nami}.
    
In the Fock representation, there must exist a particular vector 
$ |0\rangle $, called the vacuum or the no-particle state.
In a $3D$ spacetime the eigenfrequencies of the field are given by
\begin{equation}
 \omega_{n_{1}n_{2}}=\biggl(\biggl(\frac{n_{1}\pi}{L_{1}} \biggr)^{2}+
 \biggl(\frac{n_{2}\pi}{L_{2}}\biggr)^{2}\biggr)^
 {\frac{1}{2}}~~~~ n_{1},n_{2}=1,2,\dots,
\label{1}
\end{equation}
where $ L_{1} $ and $ L_{2}$ are the lengths of the sides of the box.
The zero-point energy is 
\begin{equation}
 E(L_{1},L_{2})=\frac{1}{2}\sum^{\infty}_{n_{1},n_{2}=1} 
 \omega_{n_{1}n_{2}},
\label{2}
\end{equation}
where $\omega_{n_{1}n_{2}}$ is given by Eq. (\ref{1}). This expression is 
divergent and can be written as:
\begin{equation}
 E_{\zeta}(L_{1},L_{2},s)=\frac{1}{2}\sum^{\infty}_{n_{1},n_{2}=1}
 \omega^{-2s}_{n_{1}n_{2}},
\label{3}
\end {equation}
for $s=-\frac{1}{2}$.

Eq. (\ref{3}) is analytic for $Re(s)>1$. An analytic
regularization method consists in evaluating the analytic continuation 
of the zeta function at the point $s=-\frac{1}{2}$. Algebraic manipulations 
of Eq. (\ref{3}), using Eq. (\ref{2}), give
\begin{equation}
 E_{\zeta}(L_{1},L_{2},s)=\frac{1}{8}A\biggr(\bigg(\frac{\pi}{L_{1}}\biggl)^{2},
 \biggr(\frac{\pi}{L_{2}}\biggr)^{2};s\biggr)
 -\frac{1}{4}\biggr(\biggr(\frac{L_{1}}{\pi}\biggl)^{2s}+
 \biggl(\frac{L_{2}}{\pi}\biggr)^{2s}\biggr)\zeta(2s),
\label{4}
\end{equation}
where $\zeta(2s)$ is the Riemann zeta function and $A(a,b,s)$ is the Epstein
zeta function defined as:
$$ A(a_{1},a_{2},s)=\sum^{\infty}_{n_{1},n_{2}=
-\infty}\hspace{-.25in}~'\hspace{.1in}(a_{1}n^{2}_{1}+a_{2}n^{2}_{2})^{-s}.$$
The prime sign in the summation means that the term $n_{1}=n_{2}=0$ is to be
excluded. Therefore $E_{\zeta}(L_{1},L_{2};s)$ is analytic in the complex
s-plane for $s \in C / \{\frac{1}{2}, 1\}$, and the evaluation of 
$ E_{\zeta}(L_{1},L_{2},-\frac{1}{2})$ gives 
the Casimir energy $U(L_{1},L_{2})$: 
\begin{equation}
 U(L_{1},L_{2})=\frac{\pi}{48}\biggl(\frac{1}{L_{1}}+\frac{1}{L_{2}}\biggr)-
 \frac{L_{1}L_{2}}{32\pi}\sum^{\infty}_{p,q=-\infty}
 \hspace{-.15in}~'\hspace{.1in}
 (p^{2}L^{2}_{1}+q^{2}L^{2}_{2})^{-\frac{3}{2}}.
\label{5}
\end{equation}
Instead of analytically regularizing the zero-point energy we can obtain the 
Casimir energy by introducing a suitable cut-off. For the details of these 
calculations see Ref. \cite{init}, and for a general discussion 
about analytic regularization methods used to obtain the renormalized vacuum 
energy of free fields in an arbitrary ultrastatic spacetime see Ref. 
\cite{proof}. A simple inspection of Eq. (\ref{5}) shows that the sign of the 
Casimir energy depends on the ratio between $L_{1}$ and $L_{2}$, 
and its behavior is shown in  Fig. (1) 
and Fig. (2). In the next two Sections we will extend these calculations to a 
D-dimensional spacetime assuming not only Dirichlet boundary conditions  
but also other categories of boundaries.     

\newpage

\section{ The Casimir energy of a massless scalar field 
in the presence of ``hard" boundaries in D dimensions}\

Let us consider a free massless scalar field $\varphi(t,\vec{x})$ defined 
in a $D=d+1$ dimensional Minkowski spacetime. If we assume Dirichlet b.c. 
in a D-1 dimensional box with lenghts $L_{1},L_{2},\dots,
L_{D-1}$, the eigenfrequencies are given by:
\begin{equation}
 \omega_{n_{1},n_{2},\dots,n_{D-1}}=
 \left[\left(\frac{n_{1}\pi}{L_{1}}\right)^{2}+
       \left(\frac{n_{2}\pi}{L_{2}}\right)^{2}+\dots+
       \left(\frac{n_{D-1}\pi}{L_{D-1}}\right)^{2}\right]^{1/2}.
\label{eig}
\end{equation}
Using the condition $a=L_{D-1}\ll L_{i},i=1,2,\dots,D-2$, the energy of the 
vacuum state is
\begin{equation}
 E_{D}(L_{i},a)=
 \frac{1}{2}\left(\prod_{i=1}^{D-2}L_i\right)\frac{1}{(2\pi)^{D-2}}
 \int_{0}^{\infty}dk_{1}\dots\int_{0}^{\infty}dk_{D-2}\sum_{n=1}^{\infty}
 \left[(k_{1})^{2}+\dots+(k_{D-2})^{2}+
 \left(\frac{n\pi}{a}\right)^{2}\right]^{\frac{1}{2}}.
\label{ener}
\end{equation}
Note that the summation starts at $n=1$ because for the scalar field one should
not include the modes for which all integers $n_{1},n_{2},\dots,n_{D-1}$
vanish. As was stressed in the previous Section there are two different 
ways to obtain the Casimir energy using the sum of modes method.
The first one is to use dimensional regularization in the continuous variable 
in Eq. (\ref{ener}) and analytically extend the Epstein zeta function that 
will appear after dimensional regularization. A different approach is  
to use dimensional regularization in the continuous variables and to 
introduce a cut-off in the discret one. Let us use this second approach in 
this case. For ``soft" and other types of boundaries it may be useful to 
consider both dimensional and zeta function analytic regularizations 
(see Section IV).

The angular part of the integral over the $D-2$ dimensional $k$ space can be 
calculated straightforwardly and if we define the energy per unit area by 
$\epsilon_{D}=\frac{E_{D}}{\prod L_{i}}$, for $i=1,2,\dots,D-2$, we have
\begin{equation}
 \epsilon_{D}(a)=F(D)\sum_{n=1}^{\infty}\int^{\infty}_{0}
 r^{D-3}\left(r^{2}+\left(\frac{n\pi}{a}\right)^{2}\right)^{\frac{1}{2}},
\label{den}
\end{equation}
where
\begin{equation}
 F(D)=\frac{(2\sqrt\pi)^{2-D}}{\Gamma(\frac{D-2}{2})}.
\label{f}
\end{equation}
The energy per unit area is divergent and should be regularized.
Let us introduce in Eq. (\ref{den}) a convergence factor, {\it i.e.}, an 
ultraviolet regulator
\begin{equation}
 exp\left[-\lambda\left(r^{2}+
 \left(\frac{n\pi}{a}\right)^{2}\right)^{\frac{1}{2}}\right],
\label{cut}
\end{equation}
valid for $Re(\lambda)>0$. The regularized energy per unit area is
finite provided  $Re(\lambda)>0$, and is given by
\begin{equation}
 \epsilon_{D}(\lambda,a)=F(D)\sum_{n=1}^{\infty}\int^{\infty}_{0}
 r^{D-3}\left(r^{2}+\left(\frac{n\pi}{a}\right)^{2}\right)^{\frac{1}{2}}
 exp\left[-\lambda\left(r^{2}+
 \left(\frac{n\pi}{a}\right)^{2}\right)^{\frac{1}{2}}\right].
\label{ereg}
\end{equation}
The Casimir energy per unit area (the renormalized vacuum energy per unit
area) is defined by
\begin{equation}
 U_{D}(a)=\lim\limits_{\lambda\rightarrow 0, R\rightarrow\infty}
 \left[\epsilon_{D}(\lambda,a)+\epsilon_{D}(\lambda,R-a)-
 \epsilon_{D}(\lambda,\eta R)-\epsilon_{D}(\lambda,(1-\eta)R)\right],
\label{cas1}
\end{equation}
where $\eta$ is a real number between zero and one. 
A straighforward calculation gives (see \cite{init}):
\begin{equation}
 U_{D}(a)=-\frac{1}{(4\pi)^{\frac{D}{2}}}
 \frac{\Gamma(\frac{D}{2})\zeta(D)}{a^{D-1}}.
\label{cas2}
\end{equation}
Thus the Casimir energy is negative for any $D$ in this particular 
configuration. This result is in agreement with Ambjorn and Wolfram 
\cite{ambj}. After a schematic review of this well known case, we are now 
in position to investigate two different kinds of boundaries: the ``soft" and 
``semihard" ones. It is important to stress that since we are using the sum 
of modes method to find the Casimir energy, both the cases of minimally and 
conformally coupled scalar fields are covered. As we discussed before, this 
comes from the fact that there is no difference between the density of modes 
of the minimal and of conformal scalar fields.

\section{ The effect of ``soft" and ``semihard" boundaries in the 
          Casimir energy}\

In this Section we will investigate the Casimir effect of a massless scalar 
field in the presence of ``soft" and  ``semihard" boundaries in a general 
D-dimensional spacetime. The idea is to replace the ``hard" Dirichlet walls by 
some confining potential $V(\vec{x})$ in the $x_{D-1}$ direction \cite{act}
(the Dirichlet condition corresponds to the particular case where $V(\vec{x})$
vanishes inside the cavity and $V(\vec{x})$ becomes infinite on the boundary).
In this case, the spatial modes of the scalar quantum field satisfy a
Schr\"odinger-like equation and its spectrum will be denoted by $\nu_{n}^2$.
This confining potential may be interpreted as representing some distribution
of matter with which the quantum field interacts. Since the potential acts 
as effective plates, it is relevant to state that the Casimir forces will act 
upon the matter distribution modeled by $V(\vec{x})$. When all modes are 
completely supressed only
for $x\rightarrow\infty$, the effective boundary is called ``soft". We can
also imagine an intermediate situation between ``hard" and ``soft" --- that 
may be called ``semihard" --- where the complete supression happens for a 
given finite $x$ value. In this case, the potential $V(\vec{x})$ decreases 
smoothly from an infinite value on the boundary surface $\partial\Omega$ to 
$V=0$ far from $\partial\Omega$. In this sense, Actor and Bender atribute a 
sort of ``texture" to the boundary effective surface \cite{act}, where
the case of the harmonic oscillator potential in a particular direction, 
say $x_{D-1}$, was investigated for $D=4$ (see also \cite{alb}).
 
Assuming that the boundary conditions in the direction $x_{D-1}=x$ are 
dictated by a generic potential $V(x,a)$, (where $a$ is a characteristic 
length of the system), the vacuum energy per unit area can be written as
\begin{equation}
 \epsilon_{D}(a)=-\frac{1}{2(4\pi)^{\frac{D-1}{2}}}
 \Gamma\left(\frac{1-D}{2}\right)
 \sum_{n}^{\infty}(\nu^{2}_{n})^{\frac{D-1}{2}}.
\label{pot}
\end{equation}

 The first situation that we would like to discuss is that of a potential 
which is ``semihard" near the origin and ``soft" for large $x$. 
An example of such situation can be given by the following potential, 
plotted in Fig. (3): 
\begin{equation}
 V(x,a,b)=V_{0}^{1/2}\left(\frac{a}{x}-\frac{x}{b}\right)^{2},
\label{pot1}
\end{equation}
where $a$ and $b$ have dimension of $[length]$ and $V_0$ 
has the dimension of $[length]^{-2}$. The solution of the Schr\"odinger's 
equation in the case $b=a$ is well known \cite{terHaar}. However, the above 
potential is more suitable to work out some limits and indeed only slight 
changes are needed to get the solution for $b \neq a$. It is straightforward 
to show that the energy levels of the Schr\"odinger's equation for $b\neq a$
are given by
\begin{equation}
 \nu^{2}_{n}=\left(\frac{32V_0}{b^2}\right)^{\frac{1}{2}}\left[n+\frac{1}{2}+
 \frac{1}{4}\left(\sqrt{8V_{0}a^{2}+1}-\sqrt{8V_{0}a^{2}}\right)\right],
 ~~~~ n=0,1,2,\dots
\label{esp1}
\end{equation}
Substituting Eq. (\ref{esp1}) in the vacuum energy density given by 
Eq. (\ref{pot}) we obtain:
\begin{equation}
 \epsilon_D(a,b)=-\frac{1}{2(4\pi)^{\frac{D-1}{2}}}
 \left(\frac{32V_0}{b^2}\right)^{\frac{D-1}{4}}
 \Gamma\left(\frac{1-D}{2}\right)
 \zeta\left(\frac{1-D}{2};q \right).
\label{hurw}
\end{equation}
After doing the analytic extension of the Hurwitz zeta-function 
$$\zeta(z;q)=\sum_{n=0}^{\infty}\frac{1}{(n+q)^z},$$ which is analytic at 
the beginning of an open connected set of points of the complex plane, 
{\it i.e.}, $Re(z)>1$, we obtain for the vacuum energy density: 
\begin{equation}
 \epsilon_D(a,b)=-\frac{1}{2}\frac{1}{(4\pi)^{\frac{D-1}{2}}}
 \left(\frac{32V_0}{b^2}\right)^{\frac{D-1}{4}} 
 \left[\int^{\infty}_{1}dt\,t^{-\frac{(1+D)}{2}}\,
 \frac{e^{t(1-q)}}{e^t-1}+
 \sum_{n=0}^{\infty}\frac{(-1)^{n}B_{n}(q)}{n!}
 \frac{1}{n-\frac{D+1}{2}}\right].
\label{ener1}
\end{equation}
In the above equation $B_{n}(q)$ are the Bernoulli polynomials \cite{steg} 
and $q$ is given by
\begin{equation} 
 q=\frac{1}{2}+
 \frac{1}{4}\left(\sqrt{8V_{0}a^{2}+1}-\sqrt{8V_{0}a^{2}}\right).
\label{q}
\end{equation}
To obtain the Casimir energy we have to subtract the polar part of the 
above equation, which is easily seen to be a single term in the summation, 
since the integral is finite. Two comments are in order: the first is that 
a renormalization procedure is necessary only for odd-dimensional
$(D=2m-1, m=1,2,\dots)$ spacetimes, because Eq. (\ref{hurw}) is already
finite for even $D$; second is that the ``soft" boundaries change 
the structure of the poles of the model, {\it i.e.}, the residues of the 
polar part are given by ($D$ is an odd-integer number):
 $$Res\left(\epsilon_D(a,b)\right)=
 \frac{(-1)^{\frac{D+1}{2}}}{\left(\frac{D+1}{2}\right)!}
 B_{\frac{D+1}{2}}(q).$$
Two limits are of interest (we keep $b$ fixed from now on): {\it (i)}
$a^2 \ll (V_0)^{-1}$ (hence $q\rightarrow 3/4$) in which case the potential 
behaves just as a ``hard" impenetrable (Dirichlet) wall for $x\rightarrow 0$, 
while it behaves as a harmonic oscillator potential restricted to $x>0$, and 
{\it (ii)} $a^2 \gg (V_0)^{-1}$ ($q\rightarrow 1/2$).
In the formalism of Actor and Bender \cite{act}, it is also possible to
obtain limit {\it (i)} from the harmonic oscillator (HO) result by just
discarding some of the eigenvalues in the zeta-function, and this can
be called the $\frac{1}{2}HO$ limit.

First let us investigate the limit $a^2 \ll (V_0)^{-1}$. In order to compare 
the Casimir energy with the value obtained in Ref. \cite{act}, where the 
$D=4$ case was treated, we need to use the particular value of the Hurwitz 
zeta-function 
$\zeta(-3/2;3/4)=0.02093$. In addition, one can still define 
$\lambda=(V_0/b^2)^{-1/4}$, with dimension of $[length]$, and
interpret $\lambda$ as the ``characteristic separation distance" between
the ``hard wall" at $x=0$ and the ``soft" one at $x\approx \lambda$. 
In this way, from Eq. (\ref{hurw}) we readily obtain:
\begin{equation}
 \epsilon_4(a,b)=-\frac{2^{7/4}}{3\pi}\lambda^{-3}
 \zeta(-3/2;3/4),
\label{hurw2}
\end{equation}
and $\epsilon_4$ is half the value found in Ref. \cite{act}. It should be 
noted,
however, that the value found in Ref. \cite{act} for the well known Casimir
energy in $D=4$ between two Dirichlet plates, as the limit of their result 
for the harmonic oscillator potential, is also twice the value found in
Refs. \cite{ambj,car}. In the limit of large separation between the 
``walls" ($\lambda\rightarrow\infty$), one recovers the free half-space result: 
$\epsilon\rightarrow 0^{-}$.

Let us examine now the values of the vacuum energy density given by 
Eq. (\ref{ener1}), for $D=2,3,4$, in both limiting cases $a^2 \gg (V_0)^{-1}$ 
and $a^2 \ll (V_0)^{-1}$ ($q\rightarrow 1/2$ and $q\rightarrow 3/4$ 
respectively). An interesting feature for low dimensional spacetimes, 
{\it i.e.}, for $D=2$ and $D=3$ is found: indeed, when $a^2 \gg (V_0)^{-1}$ 
the vacuum fluctuations give rise to a repulsive force corresponding to 
energy densities
\begin{equation}
 \lim_{a^2 \gg (V_0)^{-1}}\epsilon_2(a,b)=[+0.0724]\lambda^{-1}
\end{equation}
and
\begin{equation}
 \lim_{a^2 \gg (V_0)^{-1}}\epsilon_3(a,b)=[+0.0160]\lambda^{-2},
\end{equation}
respectively, while in the limit $a^2 \ll (V_0)^{-1}$ the force becomes 
attractive, corresponding to energy densities 
\begin{equation}
 \lim_{a^2 \ll (V_0)^{-1}}\epsilon_2(a,b)=[-0.0562]\lambda^{-1}
\end{equation}
and 
\begin{equation}
 \lim_{a^2 \ll (V_0)^{-1}}\epsilon_3(a,b)=[-0.0108]\lambda^{-2}.
\end{equation}
Thus, for $D=2,3$, there must exist some finite $a$ for which the Casimir 
energy vanishes. The same behavior was found in the case of the two-dimensional 
classical box considered in Section II, where the sign of the Casimir energy 
was shown to depend on the ratio between the lengths of the box \cite{cor}. 
It is important to emphasize that single poles appear in both limits only for 
$D=3$. In the four-dimensional case, for both limits, the Casimir force is 
found to be always attractive, with corresponding values for energy densities 
\begin{equation}
 \lim_{a^2 \gg (V_0)^{-1}}\epsilon_4(a,b)=[-0.0059]\lambda^{-3}
\label{a}
\end{equation}
and 
\begin{equation}
 \lim_{a^2 \ll (V_0)^{-1}}\epsilon_4(a,b)=[-0.0075]\lambda^{-3},
\label{b}
\end{equation}
and the latter is exactly the value of Eq. (\ref{hurw2}). Only in 
the limit $a^2 \ll (V_0)^{-1}$ the Casimir energy has the same (negative)
sign as in the lowest spacetime dimensions, while in the limit 
$a^2 \gg (V_0)^{-1}$ the sign of the Casimir energy for $D=4$ is opposite 
to those of $D=2$ and $D=3$. 

In Table 1 we present the values of the Casimir 
energy for a massless scalar field in the presence of the asymmetric 
potential $V(x,a,b)$ in both limits $a^2 \ll (V_0)^{-1}$ and 
$a^2 \gg (V_0)^{-1}$, for spacetime dimensionality varying in the range 
$ 2 \leq D \leq 20$. Although we went up to $D=100$, we report on the
table, for simplicity, only the values for $D \leq 20$.
From these numerical results we can discuss the sign of the Casimir energy,
which is not straightforward from the analytical expression Eq. (\ref{ener1}). 
The first point we would like to stress is that, independently of $a$, the
Casimir energy $\epsilon_D$ has the sign $(-1)^{m+1}$ for 
$D=4(m+1), m=0,1,2,3.$ For $m\geq4$, the sign becomes $(-1)^m$; indeed, for 
$D\geq18$ this pattern of signs is broken. In any case, we have checked 
that even for the wide range $2\leq D \leq 100$ the modulus of $\epsilon_D$
always decrease (we have assumed always $\lambda=1$). Secondly, we would like
to point out that, for this potential, it is possible to conceive a 
{\it gedanken experiment} to investigate the possibility of higher dimensions.
Whenever it is found, varying $a$, that the Casimir energy vanishes (and
hence changes sign), it is safe to assert that $D=4l+2$ or $D=4l+3$ ($l$
being a natural number). Alternatively, as for $D=4$, if the sign never 
changes, spacetime dimensionality should be $D=4l$ or $D=4l+1$. This is the 
maximum we can conclude from the experience in this case. But once again we 
stress that the sign pattern is altered for $D\geq 18$: while for 
$D\leq 18$ $\epsilon_{4l}$ and $\epsilon_{4l+1}$ have the same sign, for 
$D\geq 18$ they have opposite signs.

The second situation that we would like to discuss is the case of an increasing 
potential in the $x_{D-1}$ direction which becomes infinite for $x=0$ and 
$x=a$. To represent this situation let us assume that the potential is given 
by 
\begin{equation}
 V(x)=V_{0}^{1/2}\cot^{2}\left(\frac{\pi}{a}x\right).
\label{pot2}
\end{equation}
Using the Actor and Bender terminology \cite{act}, one can say that this 
situation is equivalent to two ``semihard walls", one at $x=0$, the other 
at $x=a$. Note that, in this case, the field is confined to 
the interior of the region $0<x<a$ (see Fig. (4)).  It is straightforward to 
show that the values of $\nu_{n}^{2}$ are given by
\begin{equation}
 \nu^{2}_{n}=\frac{\pi^{2}}{a^{2}}(n^{2}+4n\beta-2\beta),
 ~~~~ n=1,2,\dots
\label{esp2}
\end{equation}
where $\beta$ is defined by
\begin{equation} 
 \beta=
 \frac{1}{4}\left(\sqrt{\frac{8}{\pi^{2}}V_{0}a^{2}+1}-1\right).
\label{beta}
\end{equation}
It follows that the vacuum energy density is now given by 
\begin{equation}
 \epsilon_D(a)=-\frac{\pi^{\frac{D-1}{2}}}{2^D}
 \frac{1}{a^{D-1}}
 \Gamma\left(\frac{1-D}{2}\right)
 \sum_{n=1}^{\infty}(n^2+4n\beta -2\beta)^{\frac{D-1}{2}}.
\label{semih}
\end{equation}
Here it is necessary to analytically extend a modified Epstein zeta-function. 
This was done by Ford and also by Birrell and Ford \cite{non}. 
In this case the analytic expression for the vacuum energy density is very 
complicated and it is not reported here, but an important difference between 
the first potential and
this second one is that in some limits it is possible to recover exactly the
Dirichlet walls, {\it i.e.}, perfectly conducting plates separated by a
distance $a$. These limits can be obtained by expanding the potential around 
the point $x=\frac{a}{2}$ and neglecting terms higher than the second order. 
Let us first consider the limiting case $V_{0}\rightarrow 0$. We see that, in 
this case, we come back to the problem of classical parallel plates (Dirichlet 
b.c.) placed in the $x_{D-1}$ direction. When $V_{0}\rightarrow 0$ also 
$\beta\rightarrow 0$ and the Casimir energy reduces again to (see Eq. 
(\ref{cas2})):
\begin{equation}
 \epsilon_D(a)=-\frac{1}{(4\pi)^{\frac{D}{2}}}
 \frac{\Gamma(\frac{D}{2})\zeta(D)}{a^{D-1}},
\label{cas3}
\end{equation}
after making use of the reflection formula for $\zeta(1-D)$,
$$\Gamma(D/2)\pi^{-D/2}\zeta(D)=\Gamma\left(\frac{1-D}{2}\right)
  \pi^{\frac{D-1}{2}}\zeta(1-D).$$
For large values of the eigenvalue $\beta$, and for $n\ll\beta$, 
{\it i.e.}, for the lower levels, we get: 
\begin{equation}
 \nu^{2}_{n}=\frac{\pi}{a}(n+1/2).
\label{lambda1}
\end{equation}
This limit is completely analogous to the case of the harmonic oscilator 
potential studied in Refs. \cite{act,alb}. In the next Section we 
compute the vacuum energy of a scalar field in some configurations where
the field is constrained by different potentials in different directions.

\section{ The Casimir energy of a massless scalar field 
in a hyperbox with different boundary conditions}\

In Section IV the scalar field was supposed to be constrained by a
hyperbox where only in one direction the ``hard" Dirichlet plates were
replaced by a confining potential. In this Section we analyse different
situations in which, out of the $D$ dimensions of spacetime, in $p$ of 
them the field should satisfy Dirichlet boundary conditions, and in each 
of the remainder $D-p-1$ directions it is subjected to confining potentials.
Moreover, as we are working in Cartesian coordinates, our formalism allows 
us to choose different potentials acting upon the field in each one of the 
remainder $D-p-1$ directions.

In the directions $x^i, i=1,2,..,p$, we impose the vanishing of the
field at parallel plates located at $x^i=0,L_i, i=1,2,\dots,p$ (Dirichlet
boundary conditions); in the directions $x^j, j=p+1,p+2,\dots,D-1$, we
choose $D-p-1$ potentials $V_j(x_j,a_j)$ (all of them may be {\it a priori}
different), each one depending upon different characteristic sizes $a_j$. 
The vacuum energy is easily written as:
\begin{equation}
 E_D=\frac{1}{2}\sum_{n_1,..,n_{D-1}}\left[\left(\frac{n_1\pi}{L_1}\right)^2
 +\dots+\left(\frac{n_p\pi}{L_p}\right)^2+
 \sum_{j=p+1}^{D-1}\nu^2_{n_j}(a_j)\right]^{1/2},
\label{vac}
\end{equation}
where the functions $\nu^2_{n_j}(a_j)$ represent the spectra of eigenvalues 
of the corresponding Schr\"odinger's equation.

Now, if all $L_i$ are made much greater than all $a_j$, then we can
replace the first $p$ summations by integrals:
\begin{equation}
 E_D=\frac{1}{2}\left(\prod_{i=1}^{p}L_i\right)
 \sum_{n_{p+1},\dots,n_{D-1}}\int\frac{d^pk}{(2\pi)^p}
 \left[k^2+\sum_{j=p+1}^{D-1}\nu^2_{n_j}(a_j)\right]^{1/2},
\label{vac2}
\end{equation}
where $k$ is the length of the vector $(k_1,..,k_p)$.
The integral above is in a well-suited form to apply dimensional
regularization, with the result:
\begin{equation}
 E_D=-\frac{1}{2(4\pi)^{\frac{p+1}{2}}}\left(\prod_{i=1}^{p}L_i\right)
 \Gamma\left(-\frac{1+p}{2}\right)
 \sum_{n_{p+1},\dots,n_{D-1}}\left[\nu^2_{n_{p+1}}(a_{p+1})+\dots+
 \nu^2_{n_{D-1}}(a_{D-1})\right]^{\frac{1+p}{2}}.
\label{vac3}
\end{equation}

Instead of discussing the general case, let us work out two specific cases, 
considering a $4D$ spacetime. In one spatial direction
(for example $z$) let us impose Dirichlet boundary conditions (with plates 
separated by a distance $L$); hence $p=1$. Besides, we will subject the 
field to the same potential in the $x$ and $y$ directions:
\begin{equation}
 V(x,a,b)=V_{0}^{1/2}\left(\frac{a}{x}-\frac{x}{b}\right)^{2}
\end{equation} 
and 
\begin{equation}
 V(y,a,b)=V_{0}^{1/2}\left(\frac{a}{y}-\frac{y}{b}\right)^{2}. 
\end{equation}
The respective spectra of the Schr\"odinger's equation are 
already known to us:
\begin{equation}
 \nu^{2}_{n}=\left(\frac{32V_0}{b^2}\right)^{\frac{1}{2}}(n+q)
\end{equation}
and
\begin{equation}
 \nu^{2}_{m}=\left(\frac{32V_0}{b^2}\right)^{\frac{1}{2}}(m+q),
\end{equation}
where the value of $q$ is given by Eq. (\ref{q}). Exploiting the symmetry 
between $x$ and $y$ directions, the summation which then appears from 
Eq. (\ref{vac3}),
\begin{equation}
 \sum_{n,m=0}^{\infty}[n+m+2q],
\end{equation}
can be put in a more tractable form by using (see \cite{AActor}):
\begin{equation}
 \sum_{n,m=0}^{\infty}[n+m+c]^{-s}=\zeta(s-1;c)+(1-c)\zeta(s;c).
\label{simplif}
\end{equation} 
From the general result of Eq. (\ref{vac3}), we worked out the limits 
$a^2 \gg (V_0)^{-1}$ and $a^2 \ll (V_0)^{-1}$ (respectively $q\rightarrow 1/2$ 
and $q\rightarrow 3/4$). In the limit $a^2 \gg (V_0)^{-1}$ the vacuum energy 
is given by:
\begin{equation}
 \lim_{a^2 \gg (V_0)^{-1}} E_4(a,b)
 =L\left(\frac{V_0}{b^2}\right)^{\frac{1}{2}}\frac{\sqrt{2}}{\pi}
 \left[\int^{\infty}_{1}dt\frac{t^{-3}}{e^t-1}+
 \sum_{n=0}^{\infty}\frac{B_{n}}{n!}
 \frac{1}{n-3}\right].
\label{total} 
\end{equation}
It is now easy to obtain the energy density, {\it i.e.}, energy per unit area, 
from the expression above. First one divides the expression by $L$. 
Although the plates in the $x$ and $y$ directions are replaced by confining 
potentials, one can still assign to these directions a ``characteristic 
distance" between the ``walls", as stressed in the previous section, given by 
$\lambda=\left(\frac{V_0}{b^2}\right)^{-1/4}$. Therefore, in order to 
obtain the energy density $\epsilon$, one further divides Eq. (\ref{total}) by 
$\lambda$, which yields:
\begin{equation}
 \lim_{a^2 \gg (V_0)^{-1}} \epsilon_4(a,b)
 =\lambda^{-3}\frac{\sqrt{2}}{\pi}
 \left[\int^{\infty}_{1}dt\frac{t^{-3}}{e^t-1}+
 \sum_{n=0}^{\infty}\frac{B_{n}}{n!}
 \frac{1}{n-3}\right].
\label{lim1} 
\end{equation}
where the integral is finite and the polar part, given by the fourth term
in the summation, is identically zero because $B_3=0$ (otherwise it
would be discarded as usual). In this way, the Casimir energy in this 
limit reads:
\begin{equation}
 \lim_{a^2 \gg (V_0)^{-1}} \epsilon_4(a,b)=[-0.0069]
 \lambda^{-3}.
\label{limit1a}
\end{equation}
which is quite the same value of the limit obtained in Section IV, 
Eq. (\ref{a}) (confining potential only in one direction).

In the other limit $a^2 \ll (V_0)^{-1}$ the vacuum energy density
is given by (see Eq. (\ref{simplif})):
\begin{eqnarray}
 \lim_{a^2 \ll (V_0)^{-1}} \epsilon_4(a,b)
 &=&\lambda^{-3}\frac{\sqrt{2}}{\pi}
 \left[
 \left(\int^{\infty}_{1}dt\,
 \frac{t^{-3}e^{-t/2}}{e^t-1}+
 \sum_{n=0}^{\infty}\frac{(-1)^{n}B_{n}(3/2)}{n!}
 \frac{1}{n-3}\right) \right. \\ \nonumber
 &+&
 \left.
 \frac{1}{4}
 \left(\int^{\infty}_{1}dx\,
 \frac{x^{-2}e^{-x/2}}{e^x-1}+
 \sum_{m=0}^{\infty}\frac{(-1)^{m}B_{m}(3/2)}{m!}
 \frac{1}{m-2}\right)
 \right],
\label{lim2}
\end{eqnarray} 
where each summation contains a pole, with corresponding residues
$\frac{-B_3(3/2)}{6}=\frac{-1}{8}$ and $\frac{B_2(3/2)}{2}=\frac{11}{24}$.
The regularized vacuum energy density or, simply, the Casimir energy
in this case is evaluated to give
\begin{equation}
 \lim_{a^2 \ll (V_0)^{-1}} \epsilon_4(a,b)=[-0.0310]
 \lambda^{-3},
\label{limit1b}
\end{equation}
also negative, giving rise to an attractive Casimir force. This value is
to be compared to that of Eq. (\ref{b}).   
Thus, in any case, the replacement of two parallel Dirichlet plates in
one further direction, $y$, in comparison with the case of Section IV,
makes the absolute value of the Casimir energy to increase.

For completeness, let us calculate the Casimir energy when this ``soft"
potential acts in all three spatial directions, again for $D=4$; 
so $p=0$ in this situation. We can obtain, in this case, the vacuum energy 
density from Eq. (\ref{vac3}), by simply dividing it by 
$\lambda^2=\left(\frac{V_0}{b^2}\right)^{-1/2}$; it reads:
\begin{equation}
 \epsilon_4(a,b)=+2^{1/4}\lambda^{-3}
 \sum_{n,m,l=0}^{\infty}[n+m+l+3q]^{1/2}.
\end{equation}
Again this summation can be simplified \cite{AActor} if use is made of the
relation:
\begin{equation}
 \sum_{n,m,l=0}^{\infty}[n+m+l+c]^{-s}=\frac{1}{2}\zeta(s-2;c)
 +(\frac{3}{2}-c)\zeta(s-1;c)+\frac{1}{2}(c-1)(c-2)\zeta(s;c).
\label{simplif2}
\end{equation} 
In the limit $a^2 \gg (V_0)^{-1}$ the Casimir energy is given by:
\begin{equation}
 \lim_{a^2 \gg (V_0)^{-1}} \epsilon_4(a,b)=[-0.0132]
 \lambda^{-3},
\label{limit2a}
\end{equation}
and in the limit $a^2 \ll (V_0)^{-1}$ it reads:
\begin{equation}
 \lim_{a^2 \ll (V_0)^{-1}} \epsilon_4(a,b)=[+0.0255]
 \lambda^{-3}.
\label{limit2b}
\end{equation}
For this new configuration the result is qualitatively different from the 
situation with confining potentials only in two directions (previous case).
In the limit $a^2\gg (V_0)^{-1}$, $\epsilon_4$ is still negative and twice
the value of Eq. (\ref{limit1a}); in the other limit, $a^2\ll (V_0)^{-1}$,
$\epsilon_4$ changes sign but its absolute value is smaller than the value 
of Eq. (\ref{limit1b}).

\section{Final remarks}

We have examined how the Casimir energy of a massless scalar field confined 
to the interior of a $D-1$ dimensional hyperbox depends on different kinds of 
boundary conditions and on the dimensionality of spacetime. ``Classical" 
Dirichlet boundary conditions in one, two and three directions were relaxed; 
in these directions the constraints on the field were assumed to be given by 
some smoothly increasing
potential that represents some distribution of matter which interacts with 
the quantum field. The new contribution of the present work regards the study 
of the Casimir effect generated by two different types of boundary conditions, 
namely, the ``soft" and ``semihard" ones. In particular, we have discussed in 
details the case of a confining asymmetric potential 
$V(x)=V_0^{1/2}(a/x-x/b)^2$, which presents the feature of being ``soft" for 
$x\rightarrow\infty$ and ``semihard" for $x\rightarrow 0$.
Although the choice of the potential is in general dictated by the solvability 
of the Schr\"odinger's equation and by the manageable structure of the sum of 
proper modes, the study of how the Casimir effect depends on the boundary 
conditions opens new perspectives, which could lead to a deeper understanding 
of the interaction of real (not perfectly conducting) boundaries with the 
field.

Let us stress now some remarkable differences between the case studied here 
and the case of the Casimir energy in a $D$-dimensional ``hard" hyperbox 
considered in Ref. \cite{car}. The first one regards the repulsive or 
attractive nature of the Casimir force. In Ref. \cite{car} it was shown that 
the force is attractive if the number $p$ of finite and equal edges of a 
rectangular box is {\it odd} or for very large even values of $p$, 
irrespective of $D$. However, it was also shown in that paper that for each 
small {\it even} $p$ there exists a critical spacetime dimension $D_c(p)$ 
such that the force is repulsive if $D<D_c(p)$ and attractive otherwise. Our 
calculations have shown that, for the asymmetric potential considered here, 
there is no critical dimension $D_c$ such that for $D>D_c$ the Casimir energy 
have always the same sign. What we have demonstrated is that, independently of 
the parameter
$a$, there is a regular pattern for the sign of the Casimir energy for $D<18$; 
for other values of $D$ this sign pattern is broken (see Table 1). The 
computation of the 
vacuum energy up to $D=100$ showed that $|\epsilon_D|$ always decrease with 
increasing $D$.

A second comment concerning the sign of the Casimir energy is related to how 
it changes when $p'$ pairs of ``hard" Dirichlet walls are substituted
by ``soft" and/or ``semihard" potentials. It is well known that the ``hard" 
wall Casimir energies in $D=4$ are negative for $p=1,3$ and positive for 
$p=2$. 
A significant difference between the ``hard" and the ``soft" wall Casimir 
effects for long waveguides was first pointed out in Ref. \cite{act} and 
occurs only in the case where two pairs of 
Dirichlet plates were replaced with harmonic oscillator potential; when one 
or three pairs were substituted by the same potential, there is a qualitative 
similarity between the two cases. Our result, based on a different 
smoothly increasing potential, 
corroborates this tendency, but attention should be drawn to the following 
point when $p'=3$. Our result in the limit $a^2 \ll(V_0)^{-1}$ should be 
compared to what is called the $\frac{1}{2}HO$ limit of Ref. \cite{act}, 
which is positive and equal to our result (up to a 
systematic factor of 2, as stressed in 
the text). However, in the other asymptotic
limit $a^2 \gg (V_0)^{-1}$, the sign changes. 
Similarly, regarding ``hard" walls for $D=4$ and $p=3$, it is known that in 
the symmetric configuration with equal edges $(L_1=L_2=L_3)$, the Casimir 
energy is negative, while in Ref. \cite{cor} it was demonstrated that 
allowing $L_1 \neq L_2 \neq L_3$ the sign can also change.

As a last comment let us revisit the question of the instability of the 
semiclassical Abraham-Lorentz-Casimir model for the electron. The original 
Casimir's idea was that the electrostatic repulsion due to the electron's 
distribution of charge could be balanced by the zero-point fluctuations of 
the electromagnetic field inside and outside the conducting shell, 
assumed at that time to be ``hard".
Unfortunately, in $D=4$, the Casimir force is found to be repulsive 
\cite{boyer}. How this fact depends on $D$ was analysed in Ref. \cite{car}. 
There, it was shown that the stability of a Casimir electron model in 
higher-dimensional spacetimes would be possible only for a number of 
dimensions $D\geq8$. Therefore, one can imagine a toy model of a 
stable semiclassical electron where the Poincar\'e stress has quantum 
electromagnetic origin only if one lived in a higher-dimensional flat 
spacetime. All these results take into account that the boundaries of the 
electron are ``hard" ones. In Ref. \cite{cor} it was shown that a negative 
zero-point energy can be obtained for such b.c. only for a very 
unexpected particular 
(and antisymmetric) shape and size. Nevertheless, in the light of the new 
results for confining ``soft" boundaries, the Casimir idea of how the 
semiclassical electron could be stabilized may be revived. Indeed, there are 
two results that could be interpreted as an indication in this direction. 
First, 
it was demonstrated in Ref. \cite{act} that the Casimir energy of a spherical 
``soft" cavity is negative. Second, Eqs. (\ref{limit2a}) and (\ref{limit2b})
show that it is possible to find a particular value of $a$ which compensates 
the electrostatic repulsion in $D=4$. Thus, these results suggest how the
hypothesis of a perfect conducting shell confining the electron was
overwhelming. On the other hand, these results give 
rise to an important general question which, to the best of our knowledge, 
has no general answer yet, namely, how the sign of the Casimir energy changes 
when one changes the physical parameters and boundary conditions. It seems to 
us that the only way to discuss the attractive or repulsive character of the 
Casimir effect is, up to now, by direct computation case by case.

Finally, it should be pointed out that the mode sum energy defined by Deutsch 
and Candelas \cite{deu} stresses the fact that a classical ``perfect 
conductor" 
boundary condition is unphysical and there is a sufficiently high frequency
$\nu_0$ for which the modes are not confined by the plates (in the case of 
dieletric materials this is called the plasma frequency). In view of
this, the modes in the continuum will cancel out, leading us to assert that 
the only relevant modes to consider in the Casimir effect are the discrete 
ones. A natural extension of this paper is to consider a partially transparent 
boundary, which can be modeled, for example, by the modified P\"oschl-Teller 
potential given by $V(x)=-\frac{1}{2}\alpha\frac{\lambda(\lambda-1)}
{cosh^{2}\alpha x}$. Another direction to look on is to investigate these 
confining 
potentials in different geometries, as for example a spherical one, trying to 
generalize the results obtained by Bender and Milton \cite{sphere}.  
Interesting physical situations are those of a partially transparent sphere
and spheres with ``soft" and ``semihard" boundaries.
It is clear that this problem is of great interest in the framework of the 
bag model and may shed additional light on the Abraham-Lorentz-Casimir model
for the electron. This subject is under investigation by the authors.

\section{Acknowlegement}

 This paper was partially supported by the Conselho Nacional de 
Desenvolvimento Cient\'\i fico e Tecnologico do Brazil (CNPq).

\newpage

{\small
\begin{table} [htb]
\begin{center}
\begin{tabular} {|c||c|c|}
\hline
\hline
 $ D $ &  $a^2 \gg (V_0)^{-1}$  &  $a^2 \ll (V_0)^{-1}\,(\frac1{2}HO$ limit)  \\ 
\hline
     2    & $+0.724\times 10^{-1} \lambda^{-1}$ 
          & $-0.561\times 10^{-1} \lambda^{-1}$  \\
     3    & $+0.160\times 10^{-1} \lambda^{-2}$ 
          & $-0.108\times 10^{-1} \lambda^{-2}$  \\
     4    & $-0.590\times 10^{-2} \lambda^{-3}$ 
          & $-0.750\times 10^{-2} \lambda^{-3}$  \\
     5    & $-0.116\times 10^{-2} \lambda^{-4}$ 
          & $-0.157\times 10^{-2} \lambda^{-4}$  \\
     6    & $-0.451\times 10^{-3} \lambda^{-5}$ 
          & $+0.437\times 10^{-3} \lambda^{-5}$  \\
     7    & $-0.999\times 10^{-4} \lambda^{-6}$ 
          & $+0.823\times 10^{-4} \lambda^{-6}$  \\
     8    & $+0.335\times 10^{-4} \lambda^{-7}$ 
          & $+0.361\times 10^{-4} \lambda^{-7}$  \\
     9    & $+0.640\times 10^{-5} \lambda^{-8}$ 
          & $+0.786\times 10^{-5} \lambda^{-8}$  \\      
    10    & $+0.244\times 10^{-5} \lambda^{-9}$ 
          & $-0.243\times 10^{-5} \lambda^{-9}$  \\
    11    & $+0.541\times 10^{-6} \lambda^{-10}$ 
          & $-0.460\times 10^{-6} \lambda^{-10}$  \\
    12    & $-0.177\times 10^{-6} \lambda^{-11}$  
          & $-0.180\times 10^{-6} \lambda^{-11}$  \\
    13    & $-0.336\times 10^{-7} \lambda^{-12}$ 
          & $-0.398\times 10^{-7} \lambda^{-12}$  \\
    14    & $-0.127\times 10^{-7} \lambda^{-13}$ 
          & $+0.127\times 10^{-7} \lambda^{-13}$  \\
    15    & $-0.282\times 10^{-8} \lambda^{-14}$ 
          & $+0.241\times 10^{-8} \lambda^{-14}$  \\
    16    & $+0.914\times 10^{-9} \lambda^{-15}$ 
          & $+0.919\times 10^{-9} \lambda^{-15}$  \\
    17    & $+0.173\times 10^{-9} \lambda^{-16}$ 
          & $+0.203\times 10^{-9} \lambda^{-16}$  \\
    18    & $-0.656\times 10^{-10} \lambda^{-17}$ 
          & $+0.656\times 10^{-10} \lambda^{-17}$  \\
    19    & $+0.145\times 10^{-10} \lambda^{-18}$ 
          & $-0.124\times 10^{-10} \lambda^{-18}$  \\
    20    & $+0.470\times 10^{-11} \lambda^{-19}$ 
          & $+0.471\times 10^{-11} \lambda^{-19}$  \\

\hline
\hline
\end{tabular}
\end{center}
\caption{Some limits of the Casimir energy for confining potential 
         $V(x,a,b)=V_{0}^{1/2}\left(\frac{a}{x}-\frac{x}{b}\right)^{2}$
         in one direction and Dirichlet b.c. in the other $D-2$ directions for 
         various $D$, where $\lambda=(V_0/b^2)^{-1/4}$.}
\label{dim}
\end{table}
}

\begin{figure}[tb]
 \centerline{\epsfysize=6in\epsffile{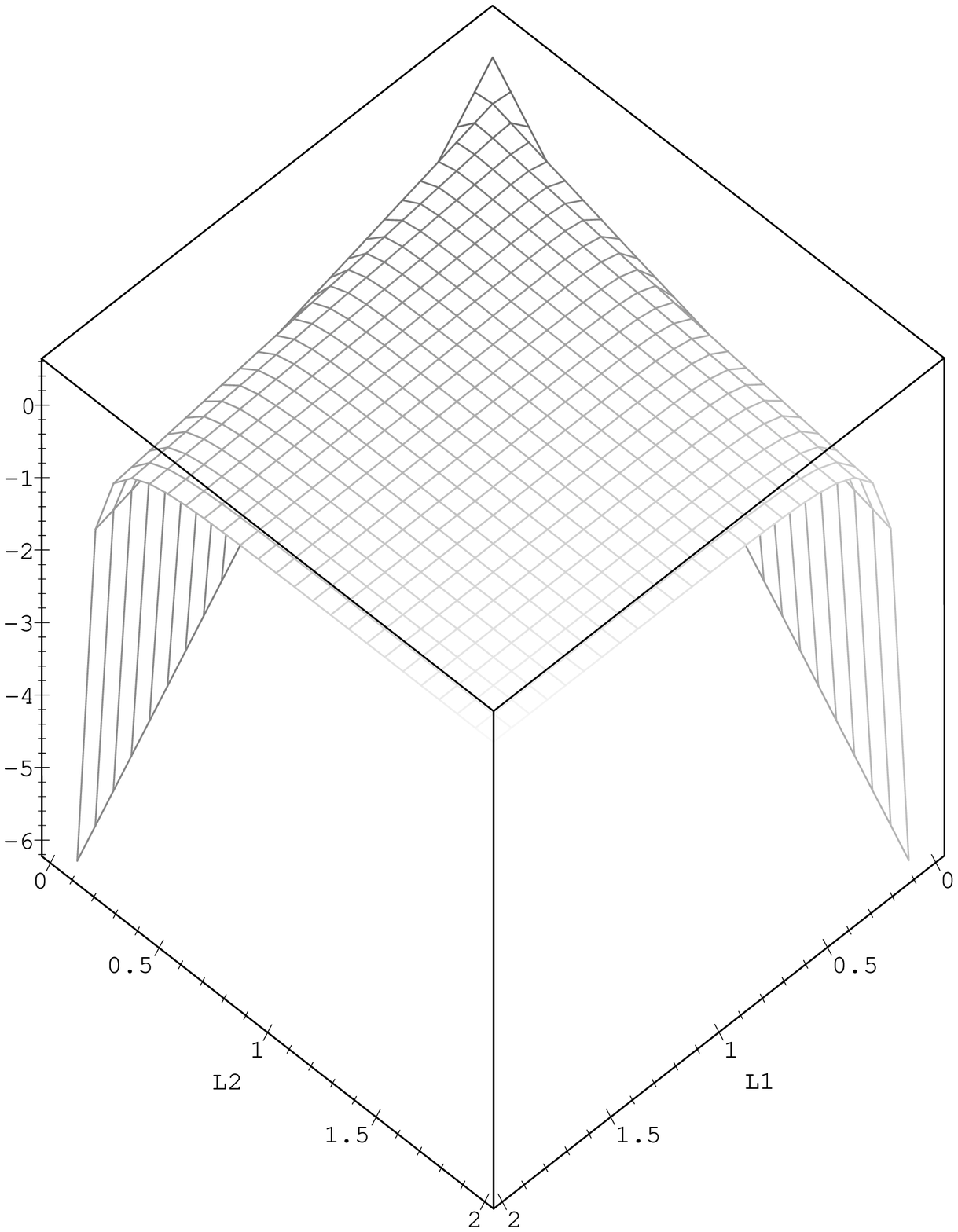}} 
 \caption[region]
 {\small\sf{ The Casimir energy of a two-dimensional classical 
             box $U(L_1,L_2)$ as a function of its lengths.}}
 \begin{picture}(10,10)
 \put(70,320){$U(L_1,L_2)$}
 
 \end{picture}
\end{figure}


\begin{figure}[tb]
 \centerline{\epsfysize=6in\epsffile{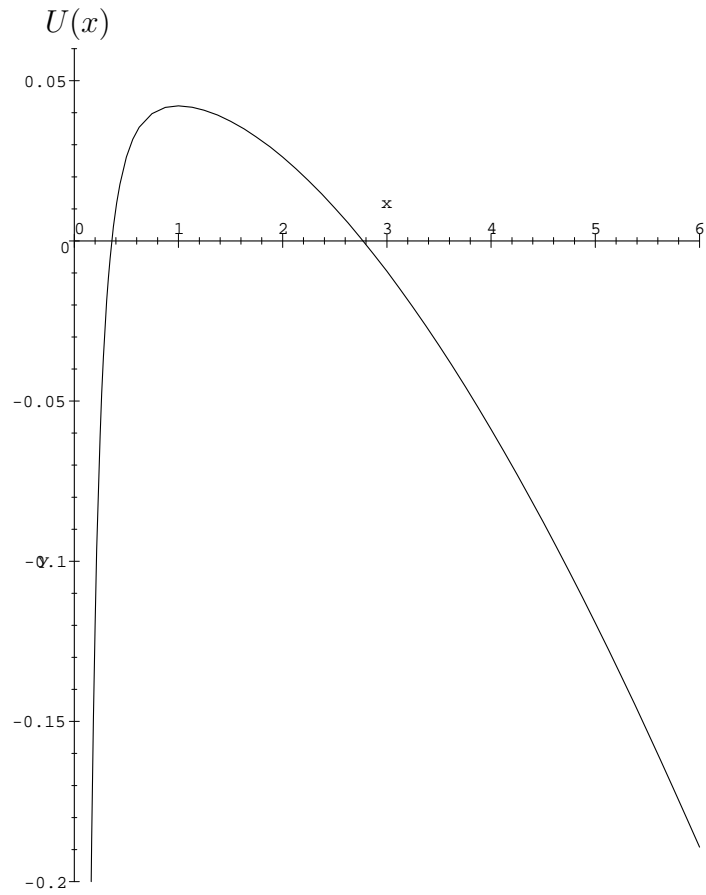}} 
 \caption[region]
 {\small\sf{ The Casimir energy of a two-dimensional classical box
             $U(L_1,L_2)$ as a function of $x=L_1/L_2$.}}
 \begin{picture}(10,10)
 \put(120,440){$U(x)$}
 
 \end{picture}
\end{figure}


\begin{figure}[tb]
 \centerline{\epsfysize=5in\epsffile{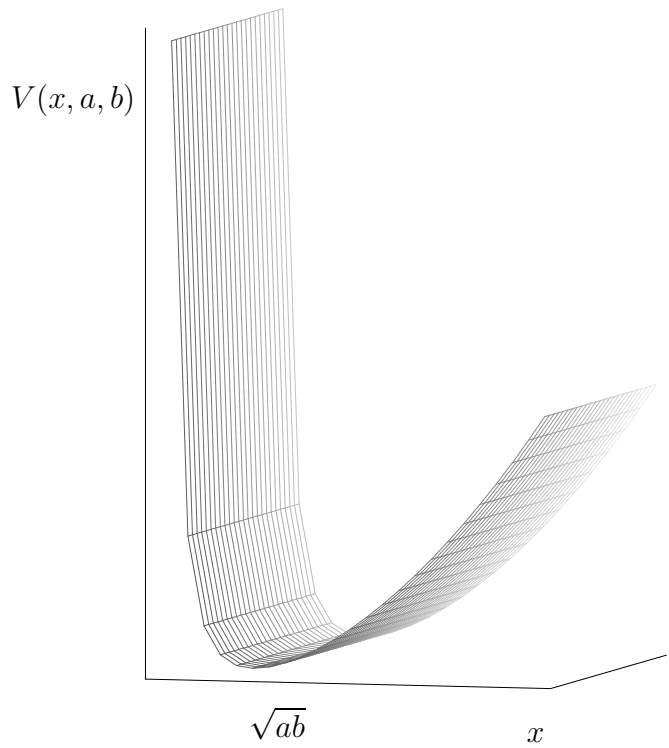}} 
 \caption[region]
 {\small\sf{  The asymmetric potential 
           $V(x,a)=V_{0}^{1/2}\left(\frac{a}{x}-\frac{x}{b}\right)^{2}$,
           which is ``semihard" for $x\rightarrow 0$ and ``soft" for 
           $x\rightarrow\infty$.}}
 \begin{picture}(10,10)
 \put(100,350){$V(x,a,b)$}
 \put(190,112){$\sqrt{ab}$}
 \put(295,110){$x$}
 
 \end{picture}
\end{figure}


\begin{figure}[tb]
 \centerline{\epsfysize=5in\epsffile{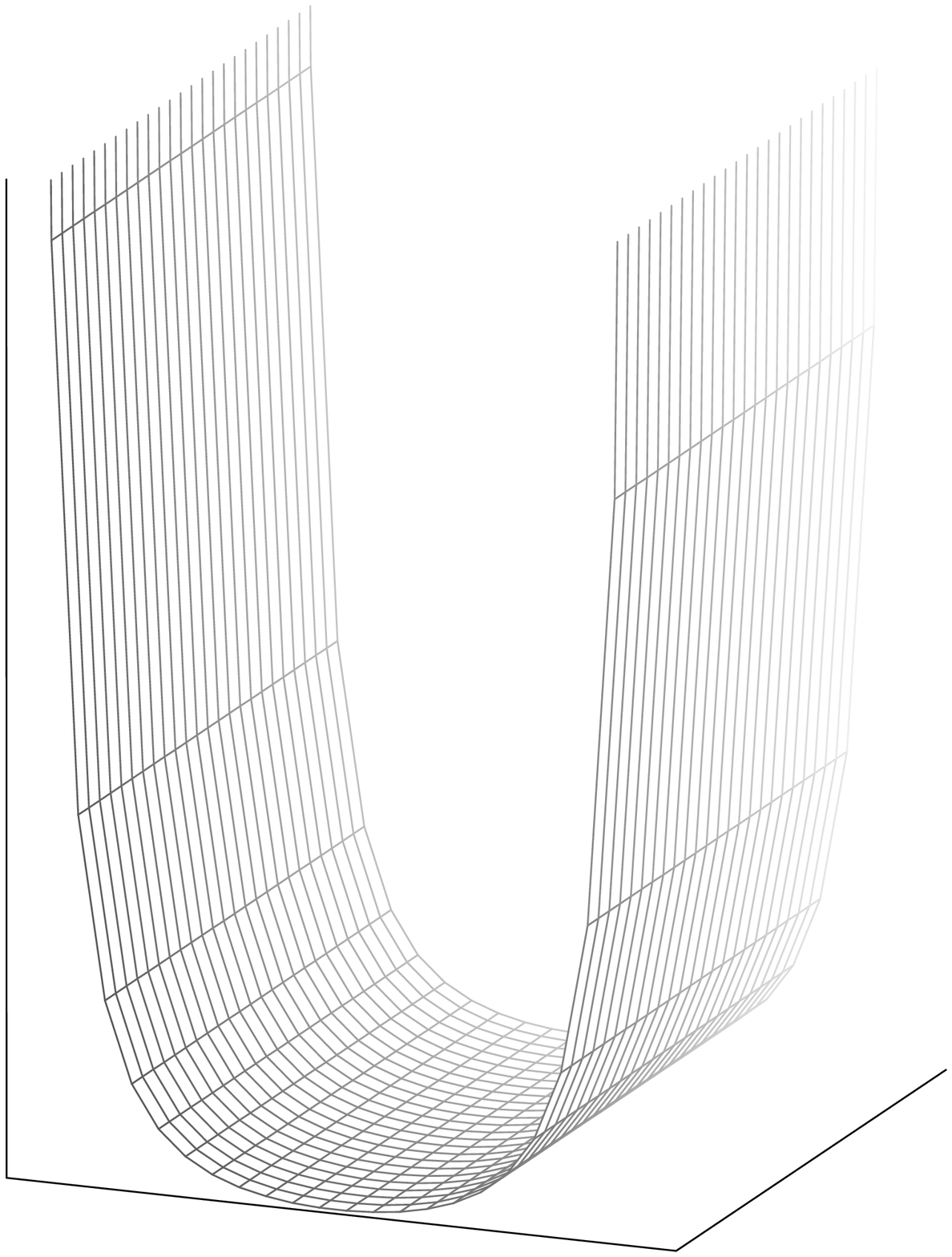}} 
 \caption[region]
 {\small\sf{  The semihard potential 
           $V(x)=V_{0}^{1/2}\cot^{2}\left(\frac{\pi}{a}x\right)$, 
           which reduces to Dirichlet walls separated by $a$ in the 
           limit $V_0\rightarrow 0$.}}
 \begin{picture}(10,10)
 \put(130,365){$V(x,a)$}
 \put(210,120){$a/2$}
 \put(285,115){$a$}

 \end{picture}
\end{figure}



\begin{thebibliography}{30}

\bibitem{act} A.A. Actor and I. Bender, {\it Phys. Rev. D} {\bf 52}, 3581 
 (1995).
\bibitem{alb} L.C. de Albuquerque, {\it Phys. Rev. D} {\bf 55}, 7754 (1997).
\bibitem{kar} H. Li and M. Kardar, {\it Phys. Rev. Lett.} {\bf 67}, 3275 
 (1991); H. Li and M. Kardar, {\it Phys. Rev. A} {\bf 46}, 6490 (1992).
\bibitem{gole} R. Golestanian and M. Kardar, "Path integral approach 
 to the dynamic Casimir effect with fluctuating boundaries", quant-ph/
 9802017; {\it idem}, "The mechanical response of vacuum", quant-ph/9701005.
\bibitem{ford} L.H. Ford and N.F. Svaiter, ``Vacuum Energy Density near
 Fluctuating Boundaries", quant-ph/9804056, CBPF pre-print CBPF-NF-007/98,
 to appear in {\it Phys. Rev. D} (1998).
\bibitem{cas} H.B.G. Casimir, {\it Proc. K. Ned. Akad. Wet.} {\bf 51},
 793 (1948); {\it Physica} {\bf 19}, 846 (1953).
\bibitem{deu} D. Deutsch and P. Candelas, 
 {\it Phys. Rev. D} {\bf 20}, 3063 (1979).
\bibitem{bro} L.H. Brown and G.J. Maclay, {\it Phys. Rev.}
 {\bf 184}, 1272 (1969).
\bibitem{DeRaad} K.A. Milton, L.L. DeRaad and J. Schwinger, {\it Ann. Phys.}
 {\bf 115}, 388 (1978).
\bibitem{cho} A. Chodos, R.L. Jaffe, K. Jonhson, C.B. Thorn and 
 W.F. Weisskopf, {\it Phys. Rev. D} {\bf 9}, 3471 (1974).
\bibitem{hays} C.M. Bender and P. Hays, {\it Phys. Rev. D}
 {\bf 14}, 2622, (1976).
\bibitem{mil1} K.A. Milton, {\it Phys. Rev. D} {\bf 22}, 1441 (1980);
 K.A. Milton, {\it Phys. Rev. D} {\bf 22}, 1444 (1980); 
 A. Chodos and C.B. Thorn, {\it Phys. Lett.} {\bf 53B}, 359 (1974).
\bibitem{rebi} D. Shalloway, {\it Phys .Rev. D} {\bf 11}, 3545 (1975); 
 C. Rebbi, {\it Phys. Rev. D} {\bf 12}, 2407 (1975).
\bibitem{creu} M. Creutz {\it Phys. Rev. D} {\bf 10}, 1749 (1974); {\it idem}, 
 {\it Phys. Rev. D} {\bf 13}, 3432 (1976). 
\bibitem{luk} W. Lukosz, {\it Physica} {\bf 56}, 109 (1971);
 S.D. Unwin, {\it Phys. Rev. D} {\bf 26}, 944 (1982).
\bibitem{ambj} J. Ambjorn and S. Wolfram, {\it Ann. Phys.} {\bf 147}, 1 (1983).
\bibitem{cor} X-z. Li, H-b. Cheng, J-m. Li and X-h. Zhai, {\it Phys. Rev. D}
 {\bf 56},  2155 (1997).
\bibitem{car} F. Caruso, N.P. Neto, B.F. Svaiter and N.F. Svaiter, 
 {\it Phys. Rev. D} {\bf 43}, 1300 (1991).
\bibitem{pet} C. Peterson, T.H. Hansson and K. Johnson, {\it Phys. Rev. D}
 {\bf 26}, 415  (1982).
\bibitem{nami} N.F. Svaiter and B.F. Svaiter, {\it J. Phys. A}
 {\bf 25}, 979 (1992).
\bibitem{init} N.F. Svaiter and B.F. Svaiter, {\it J. Math. Phys.}
 {\bf 32}, 175 (1991).
\bibitem{proof} B.F. Svaiter and N.F. Svaiter, {\it Phys. Rev. D}
 {\bf 47}, 4581 (1993); {\it idem}, {\it J. Math. Phys.} {\bf 35}, 1840 (1994).
\bibitem{terHaar} D. ter Haar, {\it Selected Problems in Quantum Mechanics},
 London, Infosearch Ltd., 1964.
\bibitem{steg} M. Abramowitz and I. Stegun (eds.),
 {\it Handbook of Mathematical Functions}, New York, Dover, 1965.
\bibitem{non} L.H. Ford, {\it Phys. Rev. D} {\bf 21}, 933 (1980);
 N.D. Birrell and L.H. Ford, {\it Phys. Rev. D} {\bf 22}, 330 (1980).
\bibitem{AActor} A.A. Actor, {\it J. Phys. A} {\bf 20}, 927 (1987).
\bibitem{boyer} T.H. Boyer, {\it Ann. Phys.} {\bf 56}, 474 (1970); B. Davies,
 {\it J. Math. Phys.} {\bf 13}, 1324 (1972).
\bibitem{sphere} C.M. Bender and K.A. Milton, {\it Phys. Rev. D} {\bf 50}, 
 6547 (1994).


\end{thebibliography}
\end{document}